\documentstyle[12pt]{article}
\begin{document}
\title{\bf The effects of nonextensive statistics on fluctuations 
investigated in event-by-event analysis of data}
\author{O.V.Utyuzh$^{1}$\thanks{e-mail: utyuzh@fuw.edu.pl},~
G.Wilk$^{1}$\thanks{e-mail: wilk@fuw.edu.pl} and Z.W\l odarczyk$^{2}$
\thanks{e-mail: wlod@pu.kielce.pl}\\[2ex] 
$^1${\it The Andrzej So\l tan Institute for Nuclear Studies}\\
    {\it Ho\.za 69; 00-689 Warsaw, Poland}\\
$^2${\it Institute of Physics, Pedagogical University}\\
    {\it  Konopnickiej 15; 25-405 Kielce, Poland}}  
\date{\today}
\maketitle

\begin{abstract}
We investigate the effect of nonextensive statistics as applied to
the chemical fluctuations in high-energy nuclear collisions discussed
recently using the event-by-event analysis of data. It turns out
that very minuite nonextensitivity changes drastically the expected
experimental output for the fluctuation measure. This results is in
agreement with similar studies of nonextensity performed recently for
the transverse momentum fluctuations in the same reactions.\\


\end{abstract}

\newpage

Some time ago a novel method of investigation of fluctuations in
even-by-event analysis of high energy multiparticle production data 
has been proposed \cite{GM}. It consists in defining a suitable
measure $\Phi$ of a given observable $x$ being exactly the same for
nucleon-nucleon and nucleus-nucleus collisions if the later are
simple superpositions of the former.  
\begin{equation}
\Phi_x\, =\, \sqrt{\frac{\left\langle Z^2\right\rangle}{\langle N\rangle}}
          \, -\, \sqrt{\bar{z^2}} 
          \qquad {\rm where}\qquad
          Z\, =\, \sum^N_{i=1}\, z_i .
           \label{eq:FI}
\end{equation}
Here $z_i = x_i - \bar{x}$ where $\bar{x}$ denotes the mean value of
the observable $x$ calculated for all particles from all events (the
so called inclusive mean) and $N$ is the number of particles analysed
in the event. In (\ref{eq:FI}) $\langle N\rangle$ and $\langle
Z^2\rangle$ are averages of event-by-event observables over all
events whereas the last term is the square root of the second
moment of the inclusive $z$ distribution. By construction \cite{GM}
if particles are produced independently $\Phi_x = 0$. \\

When applied to the recent NA49 data from central $Pb-Pb$ collisions
at $158$ A$\cdot$GeV \cite{R} this method revealed that fluctuations
of transverse momentum ($x=p_T$) decreased significantly in respect
to elementary NN collisions. This has been interpreted as a possible
sign of  equilibration taking place in heavy ion collisions providing
thus enviroment for the possible creation of quark-gluon plasma
(QGP). It was quickly realised that existing models of multiparticle
production are leading in that matter to conflicting statements
\cite{ANALYS}. On the other hand the use of fluctuations as a very
sensitive tool for the analysis of dynamics of multiparticle
reactions has been advocated since already some time \cite{GEN},
especially their role in searching for some special features of the
QGP equation of state has been shown to be of special interest
\cite{QCD}.\\  

However, it was demonstrated recently in \cite{M} that the
corresponding fluctuation measure calculated for a pion gas in global
equilibrium  (defined within the standard extensive thermodynamic) is
almost an order of magnitude greater then the experimental value.
Although the recent NA49 paper \cite{DATA} presents a new value,
which is more like the prediction in \cite{M}, the controversy
aroused around $\Phi$ resulted in a number of presentations trying to
clarify and extend  the meaning of the $\Phi$ variable (cf., for
example, \cite{OTHERS} and references therein\footnote{For some other
recent discussions of event-by-event fluctuations see
\cite{OTHERSM}.})\footnote{We would like to point here only that, if
there are some additional fluctuations (not arising from quantum
statistics, like those caused by the experimental errors) which add
in the same way to both terms in definition (\ref{eq:FI}) of $\Phi$
it would perhaps be better to use $\Phi \rightarrow \Phi^{\star} =
\frac{ \left\langle Z^2\right\rangle}{\langle N\rangle} - \bar{z^2}$
where they would cancel.}. In the mean time the use of this variable
has been extended, so far only theoretically, to the possible study
(actually  planned already by NA49) of the event-by-event
fluctuations of the "chemical" (particle type)  composition of the
final stage of high energy collisions \cite{G,MN}.\\    

Moving apart from the above discussion we would like to follow
another path of research. Namely, it was suggested recently in
\cite{FLUQ} that the extreme conditions of density and temperature
occuring in ultrarelativistic heavy ion collisions can lead to
memory effects and long-range colour interactions and to the presence
of non-Markovian processes in the corresponding kinetic equations
(cf., for example \cite{NONEQ}). It turns out that such effects in
many other branches of physics are best described phenomenologically
in terms of a single parameter $q$ by using the so called nonextensive
statistics \cite{T}. This statistics is based on new definition of
$q$-entropy (which for $q\rightarrow 1$ coincides with the usual
Boltzmann-Gibbs definition):
\begin{equation}
S_q\, =\, \frac{1}{q - 1}\, \sum^W_{k=1}\, p_k\,\left(1\, -\, p^{q-1}_k\right)
 \quad \stackrel{q \rightarrow 1}{\Rightarrow} \quad 
 S\, =\, -\sum^W_{k=1}\, p_k\, \ln p_k 
           \label{eq:S}
\end{equation}
(defined for the probability distribution $\{p_k\}$ for a system of
$W$ microstates). Such entropy is nonextensive, i.e., for a system
$(A+B)$ composed with two independent systems $A$ and $B$:
\begin{equation}
S_q(A+B) = S_q(A) + S_q(B) + (1-q)\, S_q(A)\cdot S_q(B) .\label{eq:SAB}
\end{equation}
The value of $|q-1|$ characterises deviation from the extensitivity,
see \cite{FLUQ,T} for more details. As was shown in \cite{FLUQ},
nonextensive approach with $q$ as minuite as $q=1.01\div1.015$
eliminates the abovementioned discrepancy between the first NA49 data
\cite{R} and ideal quantum gas (Boltzmann statistics) estimation of
\cite{M}.\\ 

Because NA49 Collaboration plans to study also the chemical
fluctuations, there have already apeared predictions concerning the
expected form of the fluctuation measure $\Phi$ in this case
\cite{MN}. They are based on the use of normal (i.e., Boltzmann)
statistics. In the present note we have generalized it to the
case of nonextensive statistics in a manner identical to that 
presented in \cite{FLUQ} for the case of transverse momenta. As in 
\cite{MN} we have computed the $\Phi$ measure for the system of
particles of two sorts, $\pi^-$ and $K^-$, i.e., nonstrange and
strange hadrons with multiplicities $\langle n_{\pi}\rangle$ and
$\langle n_K\rangle$, respectively. Since
\begin{equation}
\langle N\rangle \, =\, \langle n_{\pi}\rangle \, +\, 
                        \langle n_K \rangle \label{eq:PIK}
\end{equation}
one immediately finds that in definition (\ref{eq:FI})
\begin{equation}
\langle z^2\rangle\, =\, \frac{\langle n_{\pi}\rangle\,
                            \langle n_K\rangle }{\langle N\rangle}
                            \label{eq:zz}
\end{equation}
and
\begin{equation}
\langle Z^2\rangle\, =\, 
         \frac{\langle n_{\pi}  \rangle^2\, \langle n^2_K\rangle\,
           +\, \langle n^2_{\pi}\rangle\,   \langle n_K  \rangle^2
           -\, 2\, \langle n_{\pi}\rangle \langle n_K \rangle}
           {\langle N\rangle ^2} . \label{eq:ZZ}
\end{equation}
He have now consistently replaced the mean occupation numbers by
their $q$-equivalents, which under some approximations, valid for
small values of nonextensitivity $|1-q|$, can be expressed in the
following analytical form \cite{BDG}: 
\begin{equation}
\langle n\rangle_q\, =\, \left\{ \left[1\, +\,
(q-1)\beta(E-\mu)\right]^{1/(q-1)}\pm1\right\}^{-1} , \label{eq:n}
\end{equation}
where $\beta=1/kT$, $\mu$ is chemical potential and the $+/-$ sign
applies to fermions/bosons. Notice that in the limit $q \rightarrow
1$ (extensive statistics) one recovers the conventional Fermi-Dirac
and Bose-Einstein distributions. We shall not dwell on the details of
this procedure, they are essentially the same as those discussed in
\cite{FLUQ}. It is only necessary to mention that in this
approximation one retains the basic factorised formula for
correlations used in \cite{MN}, namely that 
\begin{equation}
\langle n_i n_j\rangle\, =\, \langle n_i\rangle \langle n_j\rangle .
    \label{eq:CORR}
\end{equation}    
As in \cite{FLUQ} (where $\Phi$ for transverse momenta $p_T$ has
been considered), a rather large sensitivity of predictions 
presented in \cite{MN} to the parameter $q$ has been observed.
According to the nonextensive statistics philosophy this fact
indicates a large sensitivity to the (initial and boundary)
conditions present in the ultrarelativistic heavy ion collisions and
existence of some kind of memory effects in such systems, as
mentioned in references \cite{NONEQ}. Our results are presented in
Figs. 1 and 2 where modifications caused by the nonextensitivity 
$q=1.015$ (chosen in such a way as to fit the $p_T$ spectra in Fig.
3, see discussion below) to the results of \cite{MN} for directly
produced particles are shown. For simplicity we have restricted
ourselves here only to comparison with results of \cite{MN} without
resonances \footnote{One can argue that resonance production
belongs in our philosophy already to the nonextensive case being
therefore responsible for (at least a part of) the effect leading to
a nonzero $|1-q|$. This is best seen inspecting results of \cite{MN}
with resonances included, which show that $\Phi$ in this case also
changes sign. The use of parameter $q$ is, however, more general as
it includes all other possible effects as well.}.\\   

Notice (cf. also \cite{FLUQ}) that, as is clearly seen in Fig. 3, the
same pattern of fluctuations is already present in the transverse
momentum spectra of produced secondaries, i.e., the same value
of $q$ brings new "$q$-thermal" curve in agreement with experiment in
the whole range of $p_T$ presented. If it would emerge also in the
future data on the fluctuations of chemical composition discussed
here, i.e., if parameter $q$ would turn out to be similar (modulo
experimental errors), it would signal that both observables,
fluctuations of which is investigated, are similarly affected by the
external conditions mentioned before and that they can be easily
parametrized phenomenologically by a single parameter $q$, i.e., by
the measure of the nonextensitivity of the nuclear collision
process\footnote{It is worth to mention here that methods of
nonextensive statistics have been already used in the field of high
energy physics in order  to analyse some aspects of cosmic ray data
\cite{WW} and to the  description of hadronization in $e^+e^-$
annihilation processes  \cite{BCM}. Also the recently proposed use of
quantum groups in studying Bose-Einstein correlations observed in all
multiparticle reactions \cite{BECQ} belong to that cathegory because,
as was shown in \cite{TQ}, there is close correspondence between the
deformation  parameter of quantum groups and the nonextensitivity
parameter of Tsallis statistics. In fact, as can be seen from
\cite{T}, also works on intermittency using the so called L\'evy
stable distributions (for example \cite{INTER}) belong to this
cathegory as well.}.\\

\newpage
\noindent
{\bf Figure Captions:}\\

\begin{itemize}

 \item[{\bf Fig. 1}] $\Phi$ - measure of the kaon
                     multiplicity fluctuations
                     (in the $\pi^- K^-$ system of particles)
                     as a function of temperature for three values of
                     the pion chemical potential. The kaon chemical
                     potential vanishes. The resonances are neglected. 
                     $(a)$ - results of \cite{MN} (in linear scale);
                     $(b)$ - our results for $q=1.015$. 

 \item[{\bf Fig. 2}] $\Phi$ - measure of the kaon
                     multiplicity fluctuations
                     (in the $\pi^- K^-$ system of particles)
                     as a function of temperature for three values of
                     the kaon chemical potential. The pion chemical
                     potential vanishes. The resonances are neglected. 
                     $(a)$ - results of \cite{MN} (in linear scale);
                     $(b)$ - our results for $q=1.015$. 
                     
 \item[{\bf Fig. 3}] The results for $p_T$ distribution: notice that
                     $q=1.015$ results describes also the tail of
                     distribution not fitted by the conventional
                     exponent (i.e., $q=1$ in our case, cf. also 
                     \cite{FLUQ}). Data are taken from \cite{D}.

\end{itemize}

\newpage
\begin{figure}[h]
\setlength{\unitlength}{1cm}
\begin{picture}(25.,16.5)
\includegraphics{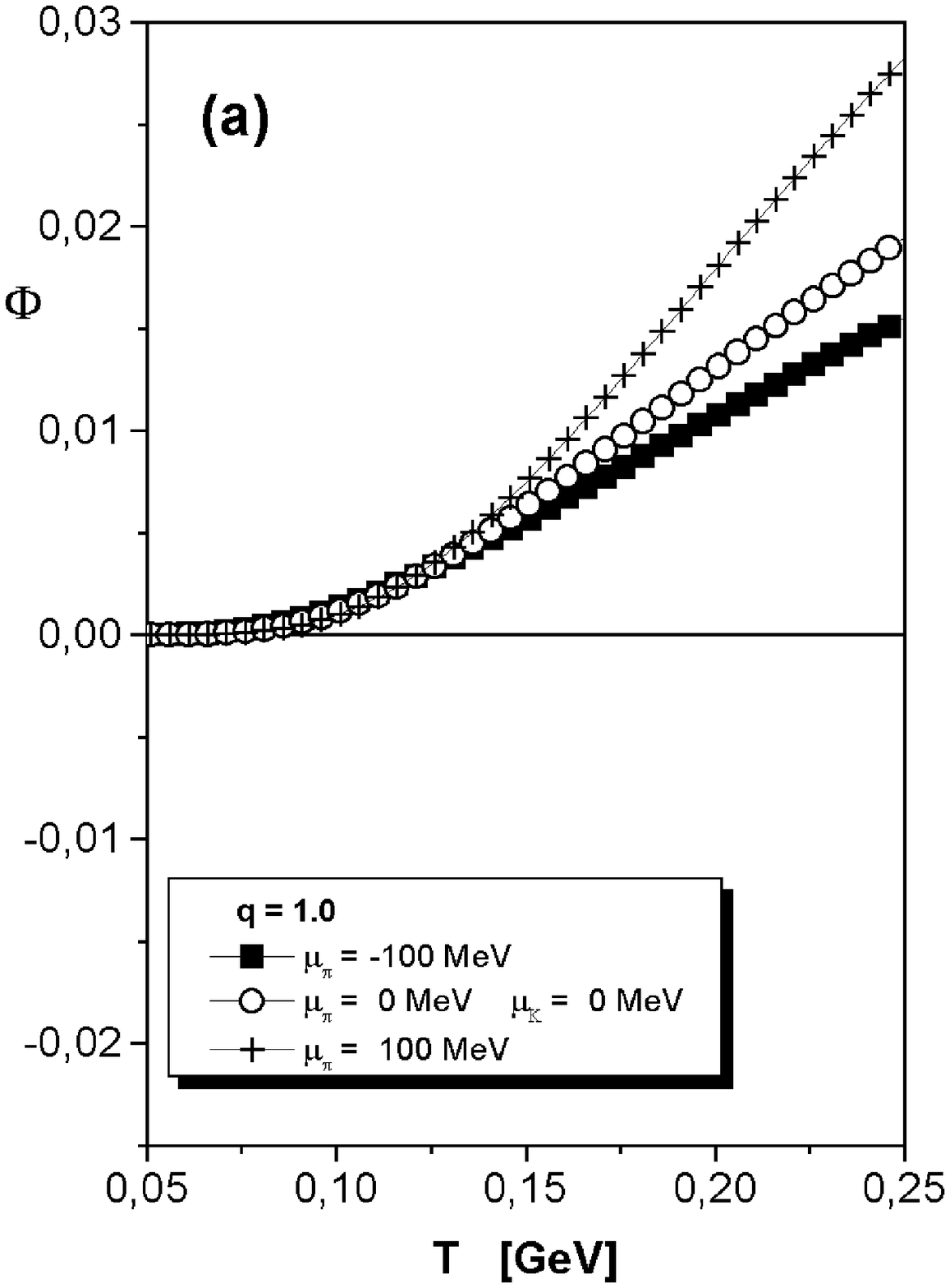}
\end{picture}
\end{figure}

\newpage
\begin{figure}[h]
\setlength{\unitlength}{1cm}
\begin{picture}(25.,16.5)
\includegraphics{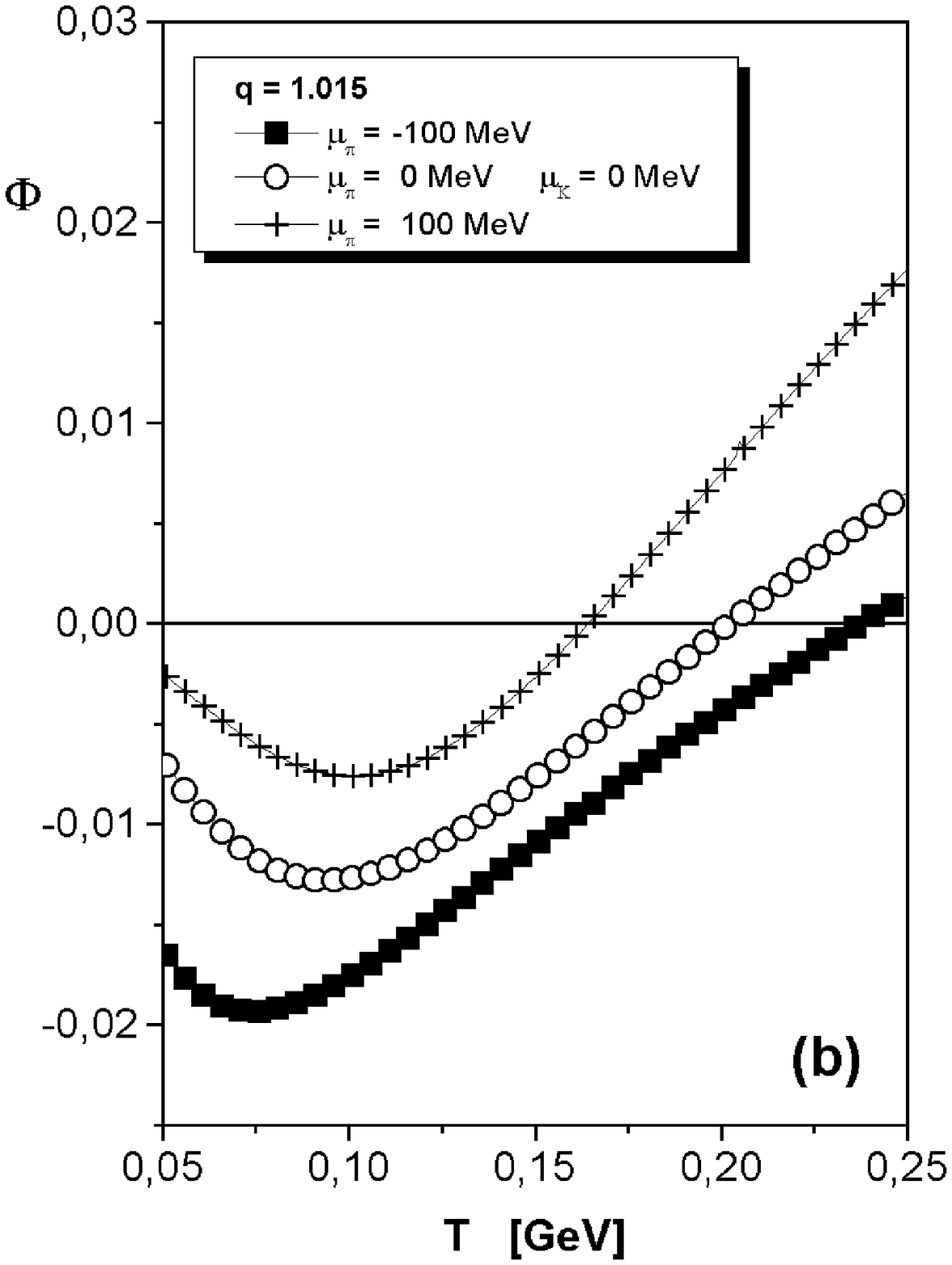}
\end{picture}
\end{figure}

\newpage
\begin{figure}[h]
\setlength{\unitlength}{1cm}
\begin{picture}(25.,16.5)
\includegraphics{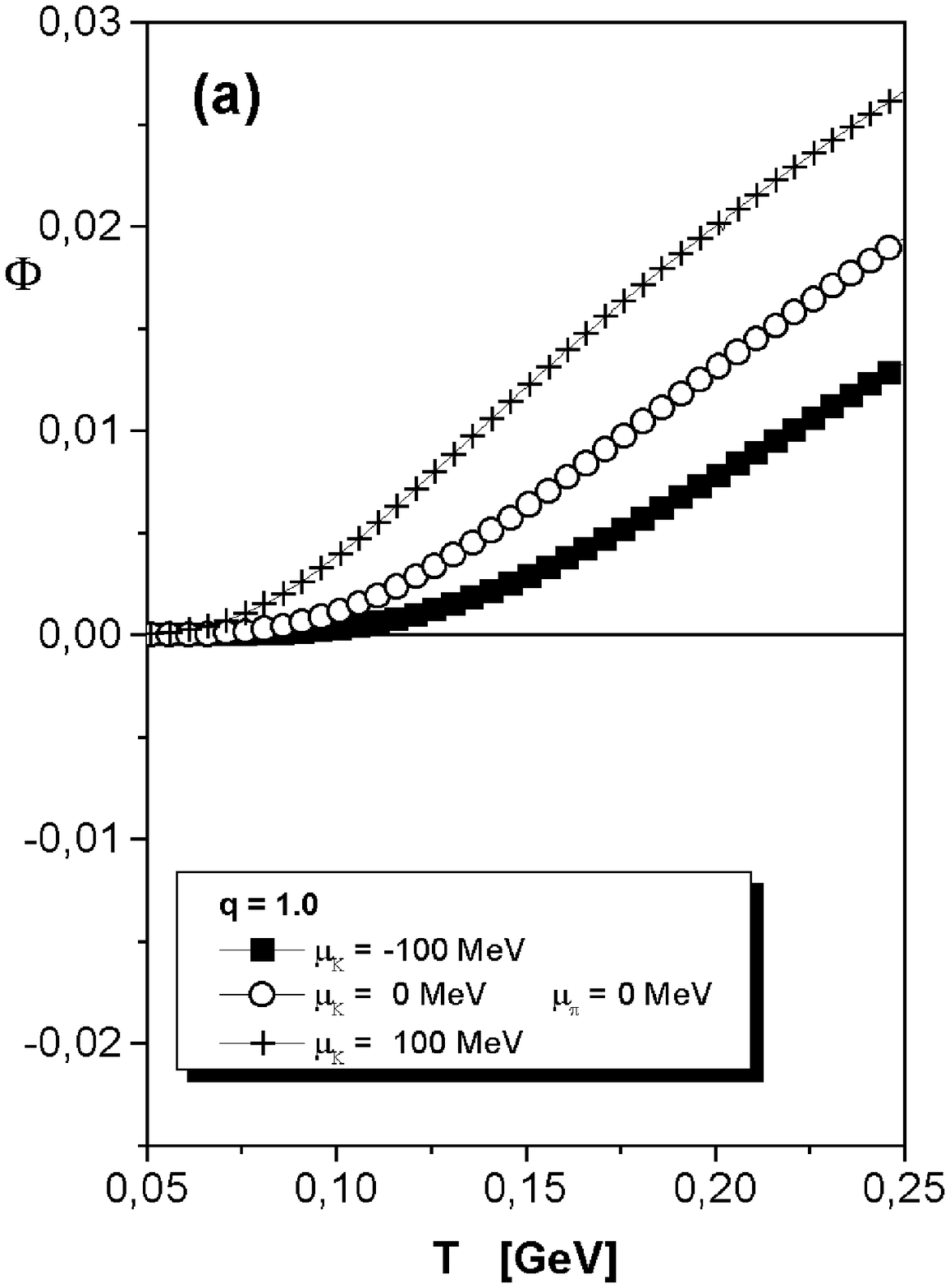}
\end{picture}
\end{figure}

\newpage
\begin{figure}[h]
\setlength{\unitlength}{1cm}
\begin{picture}(25.,16.5)
\includegraphics{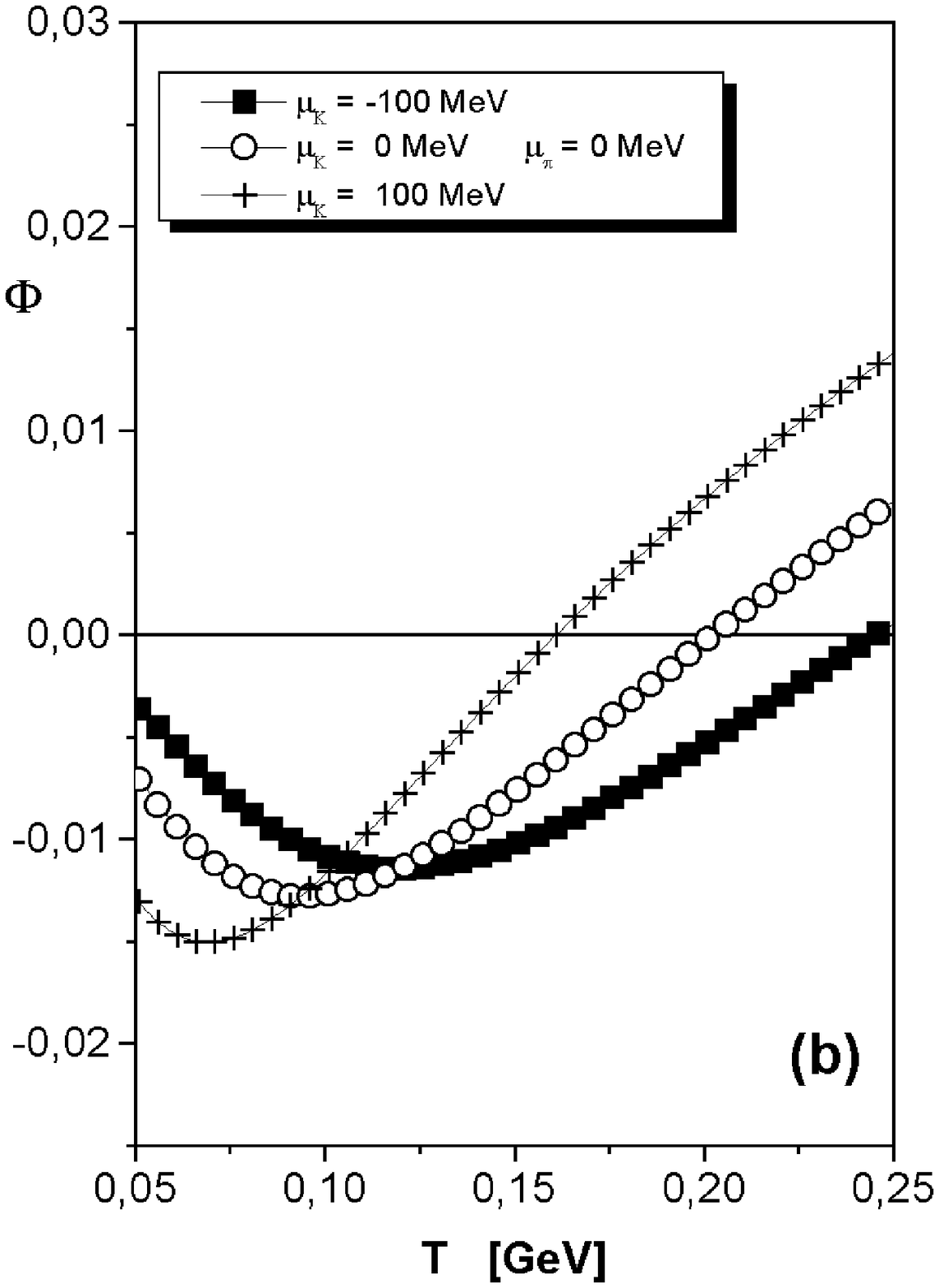}
\end{picture}
\end{figure}

\newpage
\begin{figure}[h]
\setlength{\unitlength}{1cm}
\begin{picture}(25.,16.5)
\includegraphics{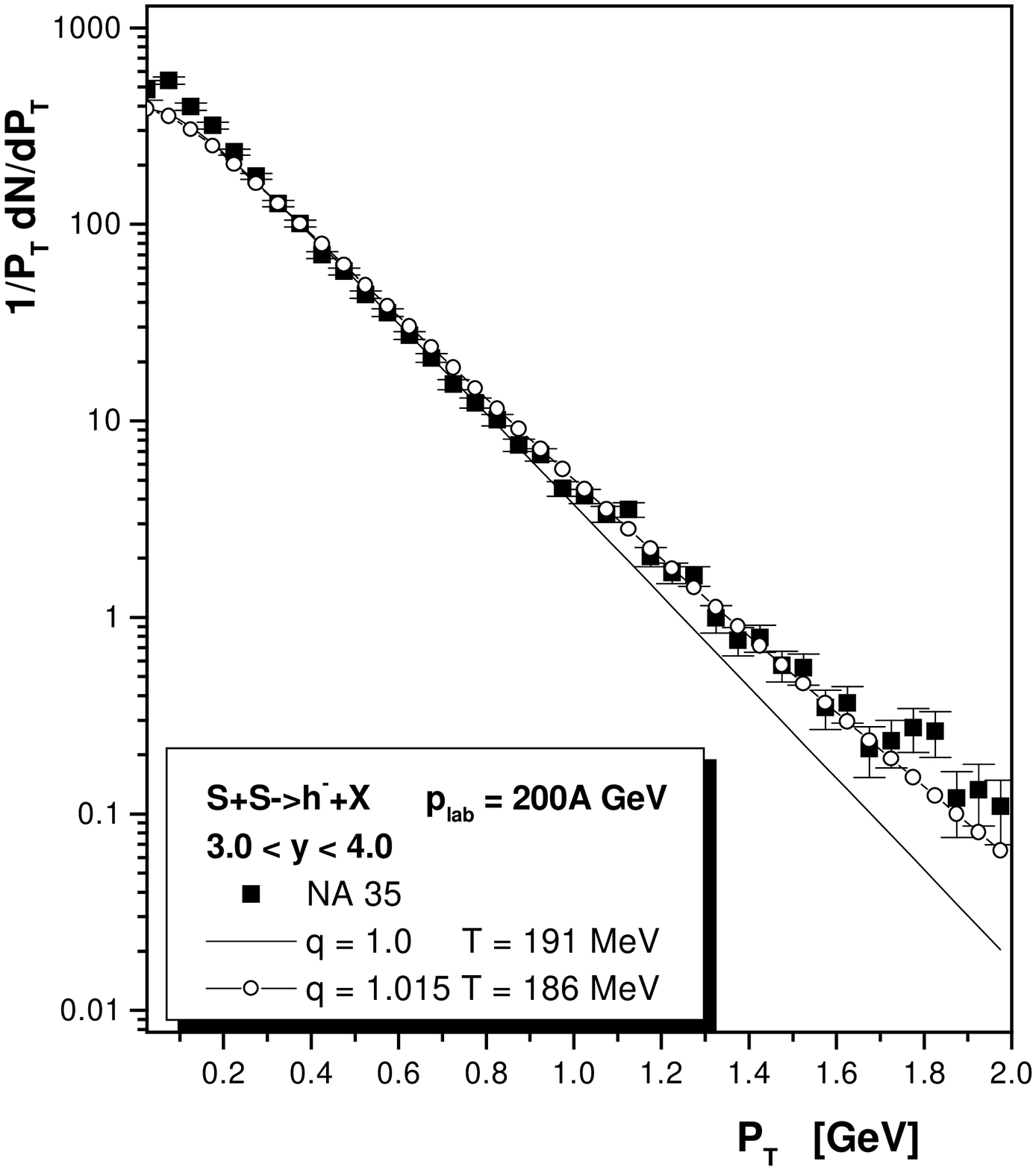}
\end{picture}
\end{figure}

\end{document}